\documentclass[runningheads]{llncs}

\usepackage{amssymb,amsmath}

\begin{document}

\pagestyle{headings}

\title{A Paraconsistent Higher Order Logic
\thanks{Originally in the proceedings of PCL 2002,
editors Hendrik Decker, J{\o}rgen Villadsen, Toshiharu Waragai
({\tt http://floc02.diku.dk/PCL/}).
Corrected.}
}

\author{J{\o}rgen Villadsen}

\institute{Computer Science, Roskilde University \\[1ex] Building 42.1, DK-4000 Roskilde, Denmark \\[2ex] \email{jv@ruc.dk}}

\maketitle

\begin{abstract}
Classical logic predicts that everything (thus nothing useful at all)
follows from inconsistency. A paraconsistent logic is a logic where an
inconsistency does not lead to such an explosion, and since in
practice consistency is difficult to achieve there are many potential
applications of paraconsistent logics in knowledge-based systems,
logical semantics of natural language, etc. Higher order logics have
the advantages of being expressive and with several automated theorem
provers available. Also the type system can be helpful. We present a
concise description of a paraconsistent higher order logic with
countable infinite indeterminacy, where each basic formula can get its
own indeterminate truth value (or as we prefer: truth code). The
meaning of the logical operators is new and rather different from
traditional many-valued logics as well as from logics based on
bilattices. The adequacy of the logic is examined by a case study in
the domain of medicine. Thus we try to build a bridge between the HOL
and MVL communities. A sequent calculus is proposed based on recent
work by Muskens.
\end{abstract}

\newcommand{\q}[1]{\ensuremath{\text{`}#1\text{'}}}

\newcommand{\KB}{\ensuremath{\mathrm{\amalg}}}

\newcommand{\V}{\mbox{$\star$}}

\newcommand{\rightqarrow}{\stackrel{\text{\tiny\sf ?}}{\rightarrow}}

\newcommand{\leftrightqarrow}{\stackrel{\text{\tiny\sf ?}}{\leftrightarrow}}

\newcommand{\Up}{\ensuremath{\mathrm{\Delta}}}

\newcommand{\Down}{\ensuremath{\nabla}}

\newcommand{\nec}{\mbox{\scriptsize $\Box$}}

\newcommand{\gen}{\mbox{\scriptsize $\partial$}}

\newcommand{\varlambda}{\mbox{\scriptsize $\lambda$}}

\newcommand{\varinfty}{\raisebox{.2ex}{\scriptsize $\infty$}}

\newcommand{\AT}{\ensuremath{\top}}
\newcommand{\AF}{\ensuremath{\bot}}
\newcommand{\AN}{\ensuremath{\dag}}
\newcommand{\AP}{\ensuremath{\ddag}}

\newcommand{\BT}{\ensuremath{\bullet}}
\newcommand{\BF}{\ensuremath{\circ}}
\newcommand{\BN}{\ensuremath{\raisebox{.15ex}{$\shortmid$}}}
\newcommand{\BP}{\ensuremath{\raisebox{.15ex}{$\shortmid\shortmid$}}}

\newcommand{\Downo}{\Down_{\!\omega}}
\newcommand{\Downi}{\Down^{\raisebox{.13ex}{$\scriptstyle \iota$}}}
\newcommand{\Downc}{\Down_{\!\!\Up}}
\newcommand{\Downci}{\Down^{\raisebox{.13ex}{$\scriptstyle \iota$}}_{\!\!\Up}}
\newcommand{\Downd}{\Down_{\!\dagger}}
\newcommand{\Downdd}{\Down_{\!\ddagger}}

\newcommand{\INT}[1]{\ensuremath{\lbrack\!\lbrack #1 \rbrack\!\rbrack}}

\makeatletter


\def\infon{\gdef\@inffactor{1}}

\def\infhide{\gdef\@inffactor{0}}

\infon

\newdimen\premisskip	
\newdimen\ursep		
\newdimen\overshoot	
\newdimen\infthick	

\newbox\pp@box		
\newbox\p@box		
\newbox\f@box		
\newbox\n@box		

\newif\if@first@infere	
\newif\if@first@and	

\newdimen\p@skip	
\newdimen\p@width	

\newdimen\pp@skip	
\newdimen\pp@end	

\premisskip	1.0em
\ursep		0.3em
\overshoot	0.3em
\infthick	0.4pt

\def\d@tmp{\dimen0}
\def\n@skip{\dimen1}
\def\f@skip{\dimen2}
\def\l@skip{\dimen3}
\def\l@width{\dimen4}

\def\@infere{
   \if@first@infere
      \@first@inferefalse
      \setbox\p@box\box\f@box
      \p@skip=0pt
      \p@width\wd\p@box
   \else
      \f@skip\p@width
      \advance\f@skip by-\wd\f@box
      \divide\f@skip by\tw@
      \advance\f@skip by\p@skip
      \ifdim\f@skip>0pt
         \n@skip=0pt
      \else
         \n@skip=-\f@skip
         \f@skip=0pt
      \fi
      \d@tmp\n@skip
      \advance\d@tmp by\p@skip
      \l@skip\mindim{\f@skip}{\d@tmp}
      \l@width\p@skip
      \advance\l@width by \p@width
      \d@tmp\f@skip
      \advance\d@tmp by \wd\f@box
      \l@width\maxdim{\l@width}{\d@tmp}
      \advance\l@width by -\l@skip
      \p@skip\f@skip
      \p@width\wd\f@box
      \setbox\p@box
      \vbox
         {\hbox{\kern\n@skip\box\p@box}
          \nointerlineskip
          \hbox{\kern\l@skip\vrule width\l@width
\raisebox{-0.5ex}[\@inffactor\infthick\infon][0pt]{\kern\ursep\box\n@box}}
          \nointerlineskip
          \hbox{\kern\f@skip\box\f@box}
         }
   \fi
}
\def\mindim#1#2{\ifdim#1<#2#1\else#2\fi}
\def\maxdim#1#2{\ifdim#1>#2#1\else#2\fi}

\def\and@proof{
   \if@first@and
      \@first@andfalse
      \pp@skip\p@skip
      \pp@end\p@skip
      \advance\pp@end by \p@width
      \setbox\pp@box\box\p@box
   \else
      \pp@end\wd\pp@box
      \advance\pp@end by\premisskip
      \advance\pp@end by\p@skip
      \advance\pp@end by\p@width
      \setbox\pp@box\hbox{\vbox{\box\pp@box}\kern\premisskip\box\p@box}
   \fi}

\def\open@f@box{\setbox\f@box\vbox\bgroup\hbox\bgroup\kern\overshoot$\strut}
\def\close@f@box{$\kern\overshoot\egroup\egroup}

\def\open@p@box{\setbox\p@box
                \vbox
                   \bgroup
                      \b@infere}
\def\close@p@box{     \e@infere
                      \global\dimen0\p@skip
                      \global\dimen1\p@width
                   \egroup
                \p@skip\dimen0
                \p@width\dimen1
}

\def\b@infere{\@first@inferetrue\open@f@box}
\def\m@infere[#1]{\close@f@box\@infere\setbox\n@box\hbox{#1}\open@f@box}
\def\e@infere{\close@f@box\@infere\box\p@box}

\def\b@and{\close@f@box\@first@andtrue\open@p@box}
\def\m@and{\close@p@box\and@proof\open@p@box}
\def\e@and[#1]{\close@p@box\and@proof
               \@first@inferefalse
               \setbox\p@box\box\pp@box
               \p@skip\pp@skip
               \p@width\pp@end
               \advance\p@width by-\p@skip
               \setbox\n@box\hbox{#1}
               \open@f@box}

\def\@@infere{\@ifnextchar [{\m@infere}{\m@infere[]}}
\def\@@and{\@ifnextchar [{\e@and}{\e@and[]}}

\newenvironment{natproof}{
   \bgroup
       \let\\=\@@infere
       \let\Lproof=\b@and
       \let\ANDproof=\m@and
       \let\Rproof=\@@and
       \b@infere
}{     \e@infere
   \egroup
}

\newenvironment{displayproof}{
    \begin{displaymath}
        \begin{natproof}
}{      \end{natproof}
    \end{displaymath}
}

\makeatother

\smallskip

\begin{small}
\begin{quote}
\it
Many non-classical logics are, at the propositional level, funny toys
which work quite good, but when one wants to extend them to higher
levels to get a real logic that would enable one to do mathematics or
other more sophisticated reasonings, sometimes dramatic troubles
appear.
\end{quote}

\noindent
J.-Y. B\'{e}ziau: \emph{The Future of Paraconsistent Logic}\\
Logical Studies Online Journal 2 (1999) p.~7
\end{small}

\newpage

\section{Introduction}

Classical logic predicts that everything (thus nothing useful at all) follows from inconsistency.
A paraconsistent logic is a logic where an inconsistency does not lead to such an explosion, and since in practice consistency
is difficult to achieve for substantial theories, paraconsistent logics have many applications in computer science, artificial intelligence, formal linguistics, etc.

In a paraconsistent logic the meaning of some of the logical operators must be different from classical logic in order to block the explosion,
and since there are many ways to change the meaning of these operators there are many different paraconsistent logics.
We present a paraconsistent higher order logic $\Down$ based on the (simply) typed $\lambda$-calculus \cite{Church40-JSL,Andrews86}.
Although it is a generalization of {\L}ukasiewicz's three-valued logic the meaning of the logical operators is new,
but with relations to logics based on bilattices \cite{Belnap77,Ginsberg88-CI,Arieli+00,Bagai98,Bagai+95-IJCM,Gottwald01}.

One advantage of a higher order logic is that the logic is very expressive in the sense that most mathematical structures, functions and relations
are available (for instance arithmetic).
Another advantage is that there are several automated theorem provers for classical higher order logic, e.g.\ HOL \cite{Gordon85}, Isabelle \cite{Paulson94},
and it should be possible to modify these to our paraconsistent logic.

We are inspired by the notion of indeterminacy as discussed by Evans \cite{Evans78-ANAL}.
Even though the higher order logic $\Down$ is paraconsistent some of its extensions, like $\Downc$, are classical.
We reuse the symbols $\Down$ and $\Up$ later for related purposes.

We also propose a sequent calculus for the paraconsistent higher order logic $\Down$ based on the seminal work by Muskens \cite{Muskens95}.
In the sequent $\Theta \vdash \Gamma$ we understand $\Theta$ as a conjunction of a set of formulas and $\Gamma$ as a disjunction of a set of formulas.
We use $\Theta \Vdash \Gamma$ as a shorthand for $\Theta,\omega \vdash \Gamma$, where $\omega$ is an axiom which provides countable infinite indeterminacy
such that each basic formula can get its own indeterminate truth value (or as we prefer: truth code).

As mentioned above higher order logic includes much, if not all, of ordinary mathematics,
and even though $\Down$ is paraconsistent we can use it for classical mathematics by keeping the truth values determinate.
Hence we shall not here consider paraconsistent mathematics.
Using the standard foundation of mathematics (axiomatic set theory) it is possible to show that $\Down$ is consistent
(but when we use the logic to build theories we might introduce inconsistencies).

The essential point is that the higher-order issues and many-valued issues complement each other in the present framework:
\begin{itemize}
\item
On the one hand we can view $\Down$ as a paraconsistent many-valued extension of classical higher order logic.
\medskip
\item
On the other hand we can view $\Down$ as a paraconsistent many-valued propositional logic with features from classical higher order logic.
\end{itemize}

First we introduce a case study in the domain of medicine and motivate our definitions of the logical operators.
Then we describe the syntax and semantics of the typed $\lambda$-calculus and introduce the primitives of the paraconsistent higher order logic $\Down$,
in particular the modality and implications available.
Finally we present a sequent calculus for $\Down$ and the extensions $\Downo$, $\Downc$, $\Downd$ and $\Downdd$.

\section{A Case Study}

As a case study we consider a small knowledge base in the domain of medicine, previously used in a very different logic programming setting \cite{Avila+97}.
Originally the knowledge base was investigated by N.~C.~A.\ da Costa and V.~S.\ Subrahmanian.
We extend the analysis developed by Villadsen \cite{Villadsen02:BALT}.

Three experts in medicine provided information related to the diagnosis of two diseases: disease-1 and disease-2.
The information concerning John and Mary can be paraphrased as follows:

\medskip

--- Expert I (a clinician):
\begin{quote}
Symptom-1 and symptom-2 together imply disease-1.

Symptom-1 and symptom-3 together imply disease-2.

Disease-1 and disease-2 exclude each other.
\end{quote}

\smallskip

--- Expert II (also a clinician):
\begin{quote}
Symptom-1 and symptom-4 together imply disease-1.

Symptom-3 implies disease-2 if symptom-1 is not present.
\end{quote}

\smallskip

--- Expert III (a pathologist):
\begin{quote}
Only John has symptom-1 and symptom-4.

Neither John nor Mary have symptom-2.

Both John and Mary have symptom-3.
\end{quote}

\smallskip

Clearly the above information is classically inconsistent, since John both has and doesn't have disease-1 and disease-2.
Hence from a straightforward formalization in classical logic we would also infer that Mary both has and doesn't have disease-1 and disease-2,
but the sensible result would be to infer just that Mary has disease-1 and doesn't have disease-2 since the inconsistency with respect to John
should not lead to inconsistency with respect to Mary.

Of course we could separate the information about John and Mary completely (in two separate knowledge bases),
but we would still be able to infer that John has, say, some other disease-3
(and doesn't have disease-3).
So we think a paraconsistent logic is needed.

It would be preferable to remove the inconsistency, but that might not be possible, either for theoretical reasons
--- what are the principles to be used in order to revise the knowledge base? ---
or for practical reasons
--- how can hundreds or thousands of evolving rules be kept consistent?
Again we think a paraconsistent logic is needed.

We would like to point out that we find the extensive literature on belief revision and update as well as
on knowledge engineering techniques a supplement rather than an alternative to works on paraconsistency.

Higher order logic is not really needed for the case study --- first order logic is enough --- but the purpose of the case study is
mainly to illustrate the working of the paraconsistency.
Even though the standard foundation of mathematics, pure axiomatic set theory, can be stated in first order logic (even as a single axiom),
higher order logics have the advantages of being expressive and with several automated theorem provers available.
Also the type system can be helpful.

We now turn to the motivation of the logical operators, which are to be defined using so-called key equalities.
We return to the case study in section \ref{cont}.

\section{Overall Motivation}

Classical logic has two truth values, namely $\BT$ and $\BF$ (truth and falsehood), and the designated truth value $\BT$ yields the logical truths.
We use the symbol $\top$ for the truth value $\BT$ and $\bot$ for $\BF$
(later these symbols are seen as abbreviations for specific formulas).

But classical logic cannot handle inconsistency since an explosion occurs.
In order to handle inconsistency we allow additional truth values and the first question is:
\begin{itemize} 
\item[1.]How many additional values do we need?
\end{itemize}

It seems reasonable to consider countably infinitely many additional truth values --- one for each proper constant we might introduce in the theory for the knowledge base.
Each proper constant (a proposition, a property or a relation) can be inconsistent ``independently'' of other proper constants.
We are inspired by the notion of indeterminacy as discussed by Evans \cite{Evans78-ANAL}.
Hence in addition to the determinate truth values $\Up = \{\bullet,\circ\}$ we also consider the indeterminate truth values
$\Down = \{\raisebox{.2ex}{$\shortmid$},\raisebox{.2ex}{$\shortmid\shortmid$},\raisebox{.2ex}{$\shortmid\shortmid\shortmid$},\ldots\}$
to be used in case of inconsistencies.
We refer to the determinate and indeterminate truth values $\Up \cup \Down$ as the truth codes.
We can then use, say,  $(\Up \cup \Down) \setminus \{\bullet\}$ as substitutes for the natural numbers $\omega = \{0,1,2,3,\ldots\}$.

The second question is:
\begin{itemize}
\item[2.]
How are we going to define the connectives?
\end{itemize}

One way to proceed is as follows.
First we want De Morgan laws to holds; hence $\varphi \lor \psi \,\equiv\, \neg (\neg \varphi \land \neg \psi)$.
For implication we have the classically acceptable $\varphi \rightarrow \psi \,\equiv\, \varphi \,\leftrightarrow\, \varphi \land \psi$.
For negation we propose to map $\BT$ to $\BF$ and vice versa, leaving the other values unchanged
(after all, we want the double negation law $\varphi \leftrightarrow \neg \neg \varphi$ to hold for all formulas $\varphi$).
For conjunction we want the idempotent law to hold and $\BT$ should to be neutral, and $\BF$ is the default result.
For biimplication we want reflexivity and $\BT$ should to be neutral, $\BF$ should be negation, and again $\BF$ is the default result.
The universal quantification is defined using the same principles as a kind of generalized conjunction
and the existential quantification follows from a generalized De Morgan law.

While it is true that $\varphi \land \neg \varphi$ does not entail arbitrary $\psi$ we do have that $\neg \varphi$ entails $\varphi \rightarrow \psi$,
hence we do not have a relevant logic \cite{Anderson+75} in general (but only for so-called first degree entailment).
Our logic validates clear ``fallacies of relevance'' like the one just noted, or like the inference from $\varphi$ to $\psi \rightarrow \psi$, but these do not
seem problematic for the applications discussed above.

Our logic is a generalization of {\L}ukasiewicz's three-valued logic, with the intermediate value duplicated many times and ordered such that
none of the copies of this value imply other ones, but it differs from {\L}ukasiewicz's many-valued logics
as well as from logics based on bilattices  \cite{Belnap77,Ginsberg88-CI,Arieli+00,Bagai98,Bagai+95-IJCM,Gottwald01}
where the third value means ``neither true nor false'' and the fourth value means ``both true and false'' (and is designated as well).

\section{Conjunction, Disjunction, and Negation}

The motivation for our logical operators is to be found in the key equalities shown to the right of the following semantic clauses
(the basic semantic clause and the clause $\INT{\AT} \,=\, \BT$ are omitted; further clauses are discussed later).
Also $\varphi \,\Leftrightarrow\, \neg \neg \varphi$ is considered to be a key equality as well.

$$
\begin{array}{ll}
\INT{\neg \varphi} ~=~ 
\left\{ \begin{array}{ll@{~~~~~~~~}c@{~~}c@{~~}c}
\BT & \text{if $\INT{\varphi} = \BF$}
&
\AT & \Leftrightarrow & \neg \AF
\\
\BF & \text{if $\INT{\varphi} = \BT$}
&
\AF & \Leftrightarrow & \neg \AT
\\
\INT{\varphi} & \text{otherwise}
&
&
&
\end{array} \right.
\end{array}
$$

\medskip

$$
\begin{array}{ll}
\INT{\varphi \land \psi} ~=~
\left\{ \begin{array}{ll@{~~~~~~~~}c@{~~}c@{~~}c}
\INT{\varphi} & \text{if $\INT{\varphi} = \INT{\psi}$}
&
\varphi & \Leftrightarrow & \varphi \land \varphi
\\
\INT{\psi} & \text{if $\INT{\varphi} = \BT$}
&
\psi & \Leftrightarrow & \AT \land \psi
\\
\INT{\varphi} & \text{if $\INT{\psi} = \BT$}
&
\varphi & \Leftrightarrow & \varphi \land \AT
\\
\BF & \text{otherwise}
&
&
&
\end{array} \right.
\end{array}
$$

\medskip

\noindent
In the semantic clauses several cases may apply if and only if they agree on the result.
The semantic clauses work for classical logic and also for our logic.

We have the following standard abbreviations:
$$
\bot \,\equiv\, \neg \top
~~~~~~~~
\varphi \lor \psi ~\equiv~ \neg (\neg \varphi \land \neg \psi)
~~~~~~~~
\exists \upsilon.\varphi \,\equiv\, \neg \forall \upsilon.\neg \varphi
$$
The universal quantification $\forall \upsilon.\varphi$ will be introduced later (as a kind of generalized conjunction).
A suitable abbreviation for $\top$ is also provided later.

In order to investigate finite truth table we first add just $\INT{\AN} = \BN$ as an indeterminacy.
We do not have $\varphi \lor \neg \varphi$.
Unfortunately we do have that $\varphi \land \neg \varphi$ entails $\psi \lor \neg \psi$
(try with $\BT$, $\BF$ and $\BN$ using the truth tables and use the fact that any $\varphi$ entails itself).
The reason for this problem is that in a sense there is not only a single indeterminacy, but a unique one for each basic formula.

However, in many situations only two indeterminacies are ever needed,
corresponding to the left and right hand side of the implication.
Hence we add $\INT{\AP} = \BP$ as the alternative indeterminacy.

\noindent
\begin{minipage}{.125\textwidth}
~
\end{minipage}%
\begin{minipage}{.25\textwidth}
\[
\begin{array}{c@{~~}c@{~~}c@{~~}c@{~~}c}
\makebox[1.2em]{$\land$} & \BT & \BF & \BN & \BP \\[.8ex]
\BT & \BT & \BF & \BN & \BP \\
\BF & \BF & \BF & \BF & \BF \\
\BN & \BN & \BF & \BN & \BF \\
\BP & \BP & \BF & \BF & \BP
\end{array}
\]
\end{minipage}%
\begin{minipage}{.25\textwidth}
\[
\begin{array}{c@{~~}c@{~~}c@{~~}c@{~~}c}
\makebox[1.2em]{$\lor$} & \BT & \BF & \BN & \BP \\[.8ex]
\BT & \BT & \BT & \BT & \BT \\
\BF & \BT & \BF & \BN & \BP \\
\BN & \BT & \BN & \BN & \BT \\
\BP & \BT & \BP & \BT & \BP
\end{array}
\]
\end{minipage}%
\begin{minipage}{.25\textwidth}
\[
\begin{array}{c@{~~}c@{~~}c@{~~}c@{~~}c}
\makebox[1.2em]{$\neg$} \\[.8ex]
\BT & \BF & ~ & ~ & ~ \\
\BF & \BT & ~ & ~ & ~ \\
\BN & \BN & ~ & ~ & ~ \\
\BP & \BP & ~ & ~ & ~
\end{array}
\]
\end{minipage}%
\bigskip

\section{Implication, Biimplication, and Modality}

As for conjunction and negation the motivation for the biimplication operator $\leftrightarrow$
(and the implication operator $\rightarrow$ as defined later) is based on the few key equalities shown to the right of the following semantic clauses.

$$
\begin{array}{ll}
\INT{\varphi \leftrightarrow \psi} ~=~
\left\{ \begin{array}{ll@{~~~~~~~~}c@{~~}c@{~~}c}
\BT & \text{if $\INT{\varphi} = \INT{\psi}$}
&
\AT & \Leftrightarrow & \varphi \leftrightarrow \varphi
\\
\INT{\psi} & \text{if $\INT{\varphi} = \BT$}
&
\psi & \Leftrightarrow & \AT \leftrightarrow \psi
\\
\INT{\varphi} & \text{if $\INT{\psi} = \BT$}
&
\varphi & \Leftrightarrow & \varphi \leftrightarrow \AT
\\
\INT{\neg \psi} & \text{if $\INT{\varphi} = \BF$}
&
\neg  \psi & \Leftrightarrow & \AF \leftrightarrow \psi
\\
\INT{\neg \varphi} & \text{if $\INT{\psi} = \BF$}
&
\neg \varphi & \Leftrightarrow & \varphi \leftrightarrow \AF
\\
\BF & \text{otherwise}
&
&
&
\end{array} \right.
\end{array}
$$

\medskip

\noindent
As before several cases may apply if and only if they agree on the result and the semantic clauses work for classical logic too.

The semantic clauses are an extension of the clauses for equality $=$:

$$
\begin{array}{ll}
\INT{\varphi = \psi} ~=~
\left\{ \begin{array}{ll@{~~~~~~~~}c@{~~}c@{~~}c}
\BT & \text{if $\INT{\varphi} = \INT{\psi}$}
\\
\BF & \text{otherwise}
\end{array} \right.
\end{array}
$$

\medskip

\noindent
We have the following abbreviations:
$$
\varphi \Leftrightarrow \psi ~\equiv~ \varphi = \psi
~~~~~~~~
\begin{array}[t]{l}
\varphi \Rightarrow \psi ~\equiv~ \varphi \,\Leftrightarrow\, \varphi \land \psi
\\[2ex]
\varphi \rightarrow \psi ~\equiv~ \varphi \,\leftrightarrow\, \varphi \land \psi
\end{array}
~~~~~~~~
\begin{array}[t]{l}
\nec \varphi \,\equiv\, \varphi = \top
\\[2ex]
\sim\! \varphi \,\equiv\, \neg \nec \varphi
\end{array}
$$
We could also have used $(\varphi \Rightarrow \psi) \land (\psi \Rightarrow \varphi)$ for $\varphi \Leftrightarrow \psi$
(using $=$ instead of $\Leftrightarrow$ in the definition of $\Rightarrow$).
Besides, $\Leftrightarrow$ binds very loosely, even more loosely than $\leftrightarrow$.

\noindent
\begin{minipage}{.125\textwidth}
~
\end{minipage}%
\begin{minipage}{.25\textwidth}
\[
\begin{array}{c@{~~}c@{~~}c@{~~}c@{~~}c}
\makebox[1.2em]{$\Leftrightarrow$} & \BT & \BF & \BN & \BP \\[.8ex]
\BT & \BT & \BF & \BF & \BF \\
\BF & \BF & \BT & \BF & \BF \\
\BN & \BF & \BF & \BT & \BF \\
\BP & \BF & \BF & \BF & \BT
\end{array}
\]
\end{minipage}%
\begin{minipage}{.25\textwidth}
\[
\begin{array}{c@{~~}c@{~~}c@{~~}c@{~~}c}
\makebox[1.2em]{$\Rightarrow$} & \BT & \BF & \BN & \BP \\[.8ex]
\BT & \BT & \BF & \BF & \BF \\
\BF & \BT & \BT & \BT & \BT \\
\BN & \BT & \BF & \BT & \BF \\
\BP & \BT & \BF & \BF & \BT
\end{array}
\]
\end{minipage}%
\begin{minipage}{.25\textwidth}
\[
\begin{array}{c@{~~}c@{~~}c@{~~}c@{~~}c}
\makebox[1.2em]{$\nec$} \\[.8ex]
\BT & \BT & ~ & ~ & ~ \\
\BF & \BF & ~ & ~ & ~ \\
\BN & \BF & ~ & ~ & ~ \\
\BP & \BF & ~ & ~ & ~
\end{array}
\]
\end{minipage}%

\noindent
\begin{minipage}{.125\textwidth}
~
\end{minipage}%
\begin{minipage}{.25\textwidth}
\[
\begin{array}{c@{~~}c@{~~}c@{~~}c@{~~}c}
\makebox[1.2em]{$\leftrightarrow$} & \BT & \BF & \BN & \BP \\[.8ex]
\BT & \BT & \BF & \BN & \BP \\
\BF & \BF & \BT & \BN & \BP \\
\BN & \BN & \BN & \BT & \BF \\
\BP & \BP & \BP & \BF & \BT
\end{array}
\]
\end{minipage}%
\begin{minipage}{.25\textwidth}
\[
\begin{array}{c@{~~}c@{~~}c@{~~}c@{~~}c}
\makebox[1.2em]{$\rightarrow$} & \BT & \BF & \BN & \BP \\[.8ex]
\BT & \BT & \BF & \BN & \BP \\
\BF & \BT & \BT & \BT & \BT \\
\BN & \BT & \BN & \BT & \BN \\
\BP & \BT & \BP & \BP & \BT
\end{array}
\]
\end{minipage}%
\begin{minipage}{.25\textwidth}
\[
\begin{array}{c@{~~}c@{~~}c@{~~}c@{~~}c}
\makebox[1.2em]{$\sim$} \\[.8ex]
\BT & \BF & ~ & ~ & ~ \\
\BF & \BT & ~ & ~ & ~ \\
\BN & \BT & ~ & ~ & ~ \\
\BP & \BT & ~ & ~ & ~
\end{array}
\]
\end{minipage}%
\bigskip

\medskip

\noindent
We could try the following standard abbreviations:
$$
\varphi \rightqarrow \psi ~\equiv~ \neg \varphi \lor \psi
~~~~~~~~
\varphi \leftrightqarrow \psi ~\equiv~ (\varphi \rightqarrow \psi) \land (\psi \rightqarrow \varphi)
$$
But here we have neither $\varphi \rightqarrow \varphi$ nor $\varphi \leftrightqarrow \varphi$, since the diagonals differ from $\BT$
at indeterminacies.

\noindent
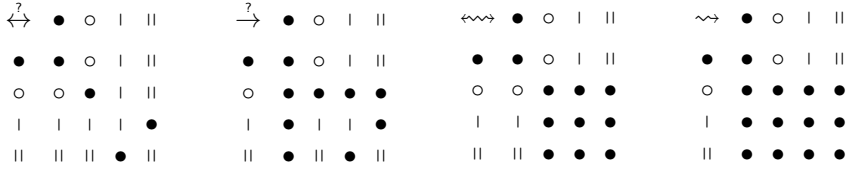
\begin{minipage}{.25\textwidth}
\[
\begin{array}{c@{~~}c@{~~}c@{~~}c@{~~}c}
\makebox[1.2em]{$\leftrightqarrow$} & \BT & \BF & \BN & \BP \\[.8ex]
\BT & \BT & \BF & \BN & \BP \\
\BF & \BF & \BT & \BN & \BP \\
\BN & \BN & \BN & \BN & \BT \\
\BP & \BP & \BP & \BT & \BP
\end{array}
\]
\end{minipage}%
\begin{minipage}{.25\textwidth}
\[
\begin{array}{c@{~~}c@{~~}c@{~~}c@{~~}c}
\makebox[1.2em]{$\rightqarrow$} & \BT & \BF & \BN & \BP \\[.8ex]
\BT & \BT & \BF & \BN & \BP \\
\BF & \BT & \BT & \BT & \BT \\
\BN & \BT & \BN & \BN & \BT \\
\BP & \BT & \BP & \BT & \BP
\end{array}
\]
\end{minipage}%
\begin{minipage}{.25\textwidth}
\[
\begin{array}{c@{~~}c@{~~}c@{~~}c@{~~}c}
\makebox[1.2em]{$\leftrightsquigarrow$} & \BT & \BF & \BN & \BP \\[.8ex]
\BT & \BT & \BF & \BN & \BP \\
\BF & \BF & \BT & \BT & \BT \\
\BN & \BN & \BT & \BT & \BT \\
\BP & \BP & \BT & \BT & \BT
\end{array}
\]
\end{minipage}%
\begin{minipage}{.25\textwidth}
\[
\begin{array}{c@{~~}c@{~~}c@{~~}c@{~~}c}
\makebox[1.2em]{$\rightsquigarrow$} & \BT & \BF & \BN & \BP \\[.8ex]
\BT & \BT & \BF & \BN & \BP \\
\BF & \BT & \BT & \BT & \BT \\
\BN & \BT & \BT & \BT & \BT \\
\BP & \BT & \BT & \BT & \BT
\end{array}
\]
\end{minipage}%
\bigskip

\medskip

\noindent
We instead use the following abbreviations:
$$
\varphi \rightsquigarrow \psi ~\equiv~ \sim\! \varphi \lor \psi
~~~~~~~~
\varphi \leftrightsquigarrow \psi ~\equiv~ (\varphi \rightsquigarrow \psi) \land (\psi \rightsquigarrow \varphi)
$$
Although $\varphi \rightsquigarrow \psi$ does not entail $\neg \psi \rightsquigarrow \neg  \varphi$,
we do have $\varphi \leftrightsquigarrow \varphi$ and $\varphi \rightsquigarrow \varphi$,
and this implication is very useful as we shall see in a moment.

We also use the predicate $\Up$ for determinacy and $\Down$ for indeterminacy with the abbreviations
(note that $\Up$ and $\Down$ are used for predicates and for sets of truth codes):
$$
\Up \varphi ~\equiv~ \nec (\varphi \lor \neg \varphi)
~~~~~~~~
\Down \varphi ~\equiv~ \neg \Up \varphi
$$

\medskip

\noindent
We now come to the central abbreviations based directly on the semantic clauses above:
$$
\varphi \leftrightarrow \psi ~\equiv~
\begin{array}[t]{l}
(\varphi = \psi \rightsquigarrow \top) ~\land \\[.3ex]
(\varphi \rightsquigarrow \psi) ~\land \\[.3ex]
(\psi \rightsquigarrow \varphi) ~\land \\[.3ex]
(\neg \varphi \rightsquigarrow \neg \psi) ~\land \\[.3ex]
(\neg \psi \rightsquigarrow \neg \varphi) ~\land \\[.3ex]
(\neg (\varphi = \psi) \land \Down \varphi \land \Down \psi \rightsquigarrow \bot)
\end{array}
$$

\medskip

\noindent
We could also use $(\varphi \leftrightsquigarrow \psi) \land (\neg \varphi \leftrightsquigarrow \neg \psi) \land
(\varphi = \psi \lor \Up \varphi \lor \Up \psi)$ for $\varphi \leftrightarrow \psi$.

\section{A Case Study --- Continued}

\label{cont}

We use the abbreviations:
$$
\varphi  \triangleright \psi ~\equiv~ \varphi \rightarrow \neg \psi
~~~~~~~~
\varphi \triangleleft\!\triangleright\, \psi ~\equiv~ (\varphi \triangleright \psi) \land (\psi \triangleright \varphi)
$$
We could also have used $\nec \neg \varphi \lor \nec \neg \psi \lor (\varphi = \psi \land \neg \nec \varphi)$
for $\varphi \triangleleft\!\triangleright\, \psi$.

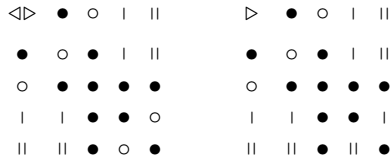
\noindent
\begin{minipage}{.25\textwidth}
~
\end{minipage}%
\begin{minipage}{.25\textwidth}
\[
\begin{array}{c@{~~}c@{~~}c@{~~}c@{~~}c}
\makebox[1.2em]{$\triangleleft\hspace*{.06em}\triangleright$} & \BT & \BF & \BN & \BP \\[.8ex]
\BT & \BF & \BT & \BN & \BP \\
\BF & \BT & \BT & \BT & \BT \\
\BN & \BN & \BT & \BT & \BF \\
\BP & \BP & \BT & \BF & \BT
\end{array}
\]
\end{minipage}%
\begin{minipage}{.25\textwidth}
\[
\begin{array}{c@{~~}c@{~~}c@{~~}c@{~~}c}
\makebox[1.2em]{$ \triangleright$} & \BT & \BF & \BN & \BP \\[.8ex]
\BT & \BF & \BT & \BN & \BP \\
\BF & \BT & \BT & \BT & \BT \\
\BN & \BN & \BT & \BT & \BN \\
\BP & \BP & \BT & \BP & \BT
\end{array}
\]
\end{minipage}%
\bigskip

\noindent
We use the operator $\triangleleft\hspace*{.06em}\triangleright$ to express the ``exclusion rule'' of expert~I.
We use the logical necessity modality operator $\nec$ to express that the observations of expert~III
concerning symptoms are not --- for the sake of simplicity --- allowed to be inconsistent.
We also use the operator $\nec$ in the ``exclusion rule'' of expert~I;
we discuss some variants later.
The operator $\nec$ is a S5 modality \cite{Hughes+68}.

A formalization is as follows with $D_i$ for disease-$i$, $S_i$ for symptom-$i$, $J$ for John and $M$ for Mary:
$$
\begin{array}{c}
S_1 x \land S_2 x ~\rightarrow~ D_1 x
~~~~~~~~
S_1 x \land S_3 x ~\rightarrow~ D_2 x
~~~~~~~~
\nec (D_1 x \triangleleft\!\triangleright\, D_2 x)
\\[2ex]
S_1 x \land S_4 x ~\rightarrow~ D_1 x
~~~~~~~~
\neg S_1 x \land S_3 x ~\rightarrow~ D_2 x
\\[2ex]
\nec S_1 J
~~~~~~~~
\nec \neg S_2 J
~~~~~~~~
\nec S_3 J
~~~~~~~~
\nec S_4 J
~~~~~~~~
\\[2ex]
\nec \neg S_1 M
~~~~~~~~
\nec \neg S_2 M
~~~~~~~~
\nec S_3 M
~~~~~~~~
\nec \neg S_4 M
~~~~~~~~
\end{array}
$$
We refer to the conjunction of these formulas as {\KB}.

We now calculate the truth code for the knowledge base {\KB}.
We do this by splitting {\KB} into {\KB}$_J$ ($x = J$ in {\KB}) and {\KB}$_M$ ($x = M$ in {\KB})
and using the truth tables we get the following two intermediate tables that must then be combined.
$$
\begin{array}{c@{~~~~~~}c@{~~~~~~}c@{~~~~~~}c@{~~~~~~~~~~}c@{~~~~~~}c@{~~~~~~}c@{~~~~~~}c}
D_1 J & D_2 J & {\KB}_J & \textrm{{\KB}'}_J & D_1 M & D_2 M & {\KB}_M & \textrm{{\KB}'}_M \\[1ex]
\BT & \BT & \BF & \BF & \BT & \BT & \BF & \BF \\
\BT & \BF & \BF & \BF & \BT & \BF & \BF & \BF \\
\BT & \BN & \BF & \BN & \BT & \BN & \BF & \BN \\
\BT & \BP & \BF & \BP & \BT & \BP & \BF & \BP \\[1ex]
\BF & \BT & \BF & \BF & \BF & \BT & \BT & \BT \\
\BF & \BF & \BF & \BF & \BF & \BF & \BF & \BF \\
\BF & \BN & \BF & \BF & \BF & \BN & \BN & \BN \\
\BF & \BP & \BF & \BF & \BF & \BP & \BP & \BP \\[1ex]
\BN & \BT & \BF & \BN & \BN & \BT & \BF & \BN \\
\BN & \BF & \BF & \BF & \BN & \BF & \BF & \BF \\
\BN & \BN & \BN & \BN & \BN & \BN & \BN & \BN \\
\BN & \BP & \BF & \BF & \BN & \BP & \BF & \BF \\[1ex]
\BP & \BT & \BF & \BP & \BP & \BT & \BF & \BP \\
\BP & \BF & \BF & \BF & \BP & \BF & \BF & \BF \\
\BP & \BN & \BF & \BF & \BP & \BN & \BF & \BF \\
\BP & \BP & \BP & \BP & \BP & \BP & \BP & \BP
\end{array}
$$

\medskip

\noindent
The columns {\KB}'$_J$ and {\KB}'$_M$ correspond to the situation where $\nec$ is omitted from the ``exclusion rule'' for expert I.

From the columns {\KB}$_J$ and {\KB}$_M$ we obtain:
$$
\begin{array}{c}
{\KB} \Vdash D_1 J
~~~~~~~~
{\KB} \Vdash \neg D_1 J
~~~~~~~~
{\KB} \Vdash \neg D_1 M
~~~~~~~~
{\KB} \nVdash D_1 M
\\[2ex]
{\KB} \Vdash D_2 J
~~~~~~~~
{\KB} \Vdash \neg D_2 J
~~~~~~~~
{\KB} \Vdash D_2 M
~~~~~~~~
{\KB} \nVdash \neg D_2 M
\end{array}
$$

\smallskip

\noindent
We consider here just the details of the first result, namely ${\KB} \Vdash D_1 J$.
For the combination {\KB} we first observe that both columns {\KB}$_J$ and {\KB}$_M$ have $\BN$ and $\BP$ rows, so {\KB} can be $\BN$ and $\BP$
({\KB} will never be $\BT$ since {\KB}$_J$ is never $\BT$, cf.\ the truth table for the conjunction operator $\land$).
We then observe that when {\KB} is $\BN$ then $D_1 J$ is $\BN$ and using the truth table for the implication operator $\rightarrow$ we get $\BT$
(the designated truth value).
Similarly for $\BP$ and hence we have ${\KB} \Vdash D_1 J$.

We find these results the best possible (given that the information is classically inconsistent)
because the inconsistency with respect to John does not lead to inconsistency with respect to Mary.

From the columns {\KB}'$_J$ and {\KB}'$_M$ we see that if $\nec$ is omitted from the ``exclusion rule'' we would not be able to derive
$\neg D_1 J$, $\neg D_1 M$ or $\neg D_2 J$.
Instead of $\triangleleft\hspace*{.06em}\triangleright$ we could consider Sheffer's stroke $|$
(where $\varphi | \psi$ is equivalent to $\neg (\varphi \land \psi)$),
but $\nec (D_1 x \,|\,  D_2 x)$ would not give any models --- all rows are $\BF$
(the $\nec$ is needed for the same reason as in the $\triangleleft\hspace*{.06em}\triangleright$ case).
However, $\nec (D_1 x \triangleleft\!\triangleright\, D_2 x ~\lor~ D_1 x \,|\, D_2 x)$ is an interesting combination where we can derive
$\neg D_1 J$ and $\neg D_2 J$, but not $\neg D_1 M$ since $D_1 M = \BN$ and $D_2 M = \BP$ give $\BP$
(and also $D_1 M = \BP$ and $D_2 M = \BN$ give $\BN$).

\section{A Sequent Calculus}

\subsection{Preliminaries}

We base the paraconsistent higher order logic $\Down$ on the (simply) typed $\lambda$-calculus \cite{Church40-JSL}
(see also \cite{Barendregt84}, especially for the untyped $\lambda$-calculus and for the notion of combinators which we use later).

Classical higher-order logic is often built from a very few primitives, say equality $=$ and the selection operator $\imath$ as in Q$_0$ \cite{Andrews86},
but it does not seem like we can avoid taking, say, negation, conjunction and universal quantification as primitives for $\Down$.
Also we prefer to extend the selection operator $\imath$ to the (global) choice operator $\varepsilon$ described later.

We use the following well-known abbreviations in order to replace negation and conjunction by joint denial (also known as Sheffer's stroke):
$$
\neg \varphi \,\equiv\, \varphi | \varphi
~~~~~~~~
\varphi \land \psi \,\equiv\, \neg (\varphi | \psi)
$$
We also have a so-called indeterminacy generation operator $\gen$ as a primitive.
We use the following abbreviations:
$$
\overline{\varphi} \,\equiv\, \neg\varphi
~~~~~~~~
\dot{\varphi} \,\equiv\, \gen \varphi ~~~~~~~~ \ddot{\varphi} \,\equiv\, \gen \dot{\varphi} ~~~~~~~~ \ldots 
$$
The indeterminacy generation operator is injective and we can use it for the natural numbers. We say much more about it later.

The truth tables in case of four truth codes are the following.

\noindent
\begin{minipage}{.125\textwidth}
~
\end{minipage}%
\begin{minipage}{.25\textwidth}
\[
\begin{array}{c@{~~}c@{~~}c@{~~}c@{~~}c}
\makebox[1.2em]{$|$} & \BT & \BF & \BN & \BP \\[.8ex]
\BT & \BF & \BT & \BN & \BP \\
\BF & \BT & \BT & \BT & \BT \\
\BN & \BN & \BT & \BN & \BT \\
\BP & \BP & \BT & \BT & \BP
\end{array}
\]
\end{minipage}%
\begin{minipage}{.25\textwidth}
\[
\begin{array}{c@{~~}c@{~~}c@{~~}c@{~~}c}
\makebox[1.2em]{$=$} & \BT & \BF & \BN & \BP \\[.8ex]
\BT & \BT & \BF & \BF & \BF \\
\BF & \BF & \BT & \BF & \BF \\
\BN & \BF & \BF & \BT & \BF \\
\BP & \BF & \BF & \BF & \BT
\end{array}
\]
\end{minipage}%
\begin{minipage}{.25\textwidth}
\[
\begin{array}{c@{~~}c@{~~}c@{~~}c@{~~}c}
\makebox[1.2em]{$\gen$} \\[.8ex]
\BT & \BT & ~ & ~ & ~ \\
\BF & \BN & ~ & ~ & ~ \\
\BN & \BP & ~ & ~ & ~ \\
\BP & \BF & ~ & ~ & ~
\end{array}
\]
\end{minipage}%
\bigskip

\medskip

\noindent
The truth table only displays equality $=$ between formulas (the biimplication operator $\Leftrightarrow$), but it is applicable to any type.
We have the abbreviation:
$$
\top \,\equiv\, (\varlambda x.x) = (\varlambda x.x)
$$
Here $\varlambda x.x$ is the identity function in the $\lambda$-calculus (any type for $x$ will do).

\subsection{Syntax}

The sets of types and terms (for each type $\tau$) are:
$$
{\cal T} \,=\, o \mid {\cal T} {\cal T} \mid {\cal S}
~~~~~~~~
{\cal L}_{\tau} \,=\, {\cal L}_{\gamma \tau} {\cal L}_{\gamma}
\mid \varlambda {\cal V}_{\alpha}.\,{\cal L}_{\beta}
\mid {\cal C}_{\tau} \mid {\cal V}_{\tau}
$$
Here $\cal S$ is the set of sorts (empty in the propositional case, where the only basic type is $o$ for formulas),
${\cal C}_{\tau}$ and ${\cal V}_{\tau}$ are the sets of term constants and variables of type $\tau$
(the set of variables must be countable infinite), and $\alpha, \beta, \gamma, \tau \in {\cal T}$
and such that $\tau = \alpha \beta$.

We often write
$\tau_{1} \ldots \tau_{m} \gamma$ instead of the type
$\tau_{1} (\ldots (\tau_{m} \gamma))$ and
$\varphi \psi_{1} \ldots \psi_{n}$ instead of the term
$((\varphi \psi_{1}) \ldots) \psi_{n}$.
Note that the relational types are $\tau_1\ldots\tau_{n}o$ (also called predicates).

If we add a sort of individuals $\iota$ to the propositional higher order logic $\Down$ we obtain the higher order logic $\Downi$
(further sorts can be added, but for our purposes they are not needed).

\subsection{Semantics}

A universe $U$ is an indexed set of type universes $U_{\tau}\neq \emptyset$ such that
$U_{\alpha \beta} \subseteq U_{\beta}^{U_{\alpha}}$.
The universe is full if $\subseteq$ is replaced by $=$.

A basic interpretation $I$ on a universe $U$ is a function
$I\colon \bigcup {\cal C}_{\tau} \to \bigcup U_{\tau}$ such that
$I \kappa_{\tau} \in U_{\tau}$ for $\kappa_\tau \in {\cal C}_\tau$.
Analogously, an assignment $A$ on a universe $U$ is a function
$A\colon \bigcup {\cal V}_{\tau} \to \bigcup U_{\tau}$ such that
$A \upsilon_{\tau} \in U_{\tau}$ for $\upsilon_\tau \in {\cal V}_\tau$.

A model $M \equiv \langle U, I \rangle$ consists of a basic
interpretation $I$ on a universe $U$ such that
for all assignments $A$ on the universe $U$
the interpretation
$\INT{\cdot}^{M,A} \colon \bigcup {\cal L}_{\tau} \to \bigcup U_{\tau}$
has $\INT{\varphi_{\tau}}^{M,A} \in U_{\tau}$ for all terms
$\varphi_{\tau} \in {\cal L}_\tau$, where
(we use the $\lambda$-calculus in the meta-language as well):
$$
\begin{array}{lll}
\INT{\kappa} & = & I \kappa \\[1ex]
\INT{\upsilon} & = & A \upsilon \\[1ex]
\INT{\varlambda \upsilon_{\alpha}.\,\varphi_{\beta}} & = &
   \varlambda u.\,\INT{\varphi}^{A [\upsilon \mapsto u]} \\[1ex]
\INT{\varphi_{\gamma \tau} \psi_{\gamma}} & = &
   \INT{\varphi} \, \INT{\psi}
\end{array}
$$
For clarity we omit some types and parameters.

What we call just a model is also known as a general model, and a full
model is then a standard model.
An arbitrary basic interpretation on a universe is sometimes considered a very general model.

\subsection{Primitives}

We use five primitive combinators of the following types ($\tau \in {\cal T}$):
$$
\begin{array}{l@{~~~~~}l@{~~~~~}l}
\textsf{D} & ooo & \text{Joint denial --- Sheffer's stroke} \\
\textsf{Q} & \tau\tau o & \text{Equality} \\
\textsf{A} & (\tau o)o & \text{Universal quantification} \\
\textsf{C} & (\tau o)\tau & \text{Global choice} \\
\textsf{V} & oo & \text{Indeterminacy generation}
\end{array}
$$
We have the following abbreviations (we omit the types):
$$
\varphi | \psi \,\equiv\, \textsf{D}\varphi\psi
~~~~~~~~
\varphi = \psi \,\equiv\, \textsf{Q}\varphi\psi
~~~~~~~~
\breve{\varphi} \,\equiv\, \textsf{Q}\varphi
$$
$$
\forall \upsilon.\varphi \,\equiv\, \textsf{A}\, \varlambda \upsilon.\varphi
~~~~~~~~
\tilde{\varphi} \,\equiv\, \textsf{C}\varphi
~~~~~~~~
\varepsilon \upsilon.\varphi \,\equiv\, \textsf{C}\, \varlambda \upsilon.\varphi
~~~~~~~~
\gen \varphi \,\equiv\, \textsf{V}\varphi
$$
Only a few of these abbreviations need explanation.
The (global) choice operator $\varepsilon$ chooses \emph{some} value $x$ for which $\varphi$ is satisfied ($x$ can be free in $\varphi$);
if no such value $x$ exists then an arbitrary value is chosen (of the right type).
The choice is global in the sense that all choices are the same for equivalent $\varphi$'s, hence for instance we have $(\varepsilon x.\bot) = (\varepsilon x.\bot)$.

The notation $\breve{\varphi}$ turns $\varphi$ into a singleton set with itself as the sole member and $\tilde{\varphi}$ is its inverse,
since $\tilde{\breve{\varphi}} = \varphi$, which is called the selection property, cf.\ the selection operator $\imath$ in Q$_0$ \cite{Andrews86}.
But $\tilde{\varphi}$ is of course also defined for non-singleton sets, namely as the (global) choice operator just described.
We say a little more about these matters when we come to the choice rules.

We can even eliminate the $\lambda$-notation if we use two additional primitive combinators, the so-called \textsf{S} and \textsf{K} combinators of suitable types.
For example, the identity function $\varlambda x.x$ is available as the abbreviation $\textsf{I} \,\equiv\, \textsf{S} \textsf{K} \textsf{K}$, cf.\ \cite{Barendregt84}.

\subsection{Structural Rules}

In the sequent $\Theta \vdash \Gamma$ we understand $\Theta$ as a conjunction of a set of formulas and $\Gamma$ as a disjunction of a set of formulas,
and we have the usual rules for a monotonic sequent calculus:

$$
\begin{natproof}
\Lproof
\Theta,\varphi ~\vdash~ \Gamma
\ANDproof
\Theta ~\vdash~ \varphi,\Gamma
\Rproof[Cut]
\Theta ~\vdash~ \Gamma
\end{natproof}
~~~~~~~~
\begin{natproof}
\Theta ~\vdash~ \Gamma
\\
\Theta,\varphi ~\vdash~ \Gamma
\end{natproof}
~~~~~~~~
\begin{natproof}
\Theta ~\vdash~ \Gamma
\\
\Theta ~\vdash~ \varphi,\Gamma
\end{natproof}
$$

\medskip

\noindent
Notice that $\varphi ~\vdash~ \varphi$ follows from these rules and the rules for equality below.

\subsection{Fundamental Rules}

We use the abbreviation:
$$
\varphi \stackrel{\forall}{=} \psi \,\equiv\, (\varlambda p q.\, \forall x.\, p x = q x) \varphi \psi
$$
We have the usual conversion and extensionality axioms of the $\lambda$-calculus:
$$
(\varlambda \upsilon.\varphi)\,\psi = \varphi[\psi/\upsilon]
~~~~~~~~
\varphi \stackrel{\forall}{=} \psi ~\vdash~ \varphi = \psi
$$
Here $\varphi[\psi/\upsilon]$ means the substitution of $\psi$ for the variable $\upsilon$ in $\varphi$
(the notation presupposes that $\psi$ is substitutable for $\upsilon$ in $\varphi$).
For later use we note that if the notation for an arbitrary so-called eigen-variable $\pi$
is used in place of $\psi$ then it must not occur free in other formulas in the given axiom/rule).
Also $\varphi[\psi]$ means $\varphi[\psi/\upsilon]$ for an arbitrary variable $\upsilon$ with respect to the given axioms/rule.

We have the usual reflexivity and substitution axioms for equality:
$$
\varphi = \varphi
~~~~~~~~
\varphi = \psi,~ \theta[\varphi] ~\vdash~ \theta[\psi]
$$

\subsection{Logical Rules}

Let $\overline{\Theta} = \{\overline{\theta} \mid \theta \in \Theta\}$.
Negation is different from classical logic.
We follow \cite{Muskens95} and add only the following rules:

$$
\begin{natproof}
\overline{\Gamma} ~\vdash~ \Theta
\\
\overline{\Theta} ~\vdash~ \Gamma
\raisebox{0pt}[2.5ex]{}
\end{natproof}
~~~~~~~~
\begin{natproof}
\Gamma ~\vdash~ \overline{\Theta}
\\
\Theta ~\vdash~ \overline{\Gamma}
\raisebox{0pt}[2.5ex]{}
\end{natproof}
$$

\medskip

\noindent
Conjunction and universal quantification are straightforward:
$$
\raisebox{1.25ex}{$\varphi,\psi ~\vdash~ \varphi \land \psi$}
~~~~~~~~
\begin{natproof}
\Theta, \varphi,\psi ~\vdash~ \Gamma
\\
\Theta, \varphi \land \psi ~\vdash~ \Gamma
\end{natproof}
$$

$$
\raisebox{1.25ex}{$\forall \upsilon.\varphi ~\vdash~ \varphi[\psi/\upsilon]$}
~~~~~~~~
\begin{natproof}
\Theta ~\vdash~ \varphi[\pi/\upsilon],\Gamma
\\
\Theta ~\vdash~ \forall \upsilon.\varphi,\Gamma
\end{natproof}
$$

\medskip

\noindent
Remember that the eigen-variable condition is built into the notation.

We also have to provide axioms for the negation and conjunction in case of indeterminacy:
$$
\Down x \,\rightsquigarrow\, \neg x = x
~~~~~~~~
x \neq y \land \Down x \land \Down y \,\rightsquigarrow\, x | y
$$

\subsection{Choice Rules}

We have the following choice axioms \cite{Church40-JSL} corresponding to the Axiom of Choice in axiomatic set theory:
$$
p \upsilon \rightsquigarrow p \tilde{p}
$$

\medskip

\noindent
Notice that due to the use of $\rightsquigarrow$ we can only make a choice if $\exists \upsilon.\, \nec(p \upsilon)$.
If we used a different implication the choice might not be possible at all.

\subsection{Generation Rules}

We use the following abbreviations:
$$
\varinfty \,\equiv\, \top
~~~~~~~~
0 \,\equiv\, \bot
~~~~~~~~
1 \,\equiv\, \dot{0}
~~~~~~~~
2 \,\equiv\, \dot{1}
~~~~~~~~
3 \,\equiv\, \dot{2} 
~~~~~~~~
\ldots
$$
$$
\mathbb{N} \,\equiv\, \varlambda x.\, x \neq \varinfty
~~~~~~~~
\mathbb{T} \,\equiv\, \varlambda x.\top
~~~~~~~~
\emptyset \,\equiv\, \varlambda x.\bot
$$

\medskip

\noindent
We have the following important axioms:
$$
\dot{x} = \dot{y} \,\rightsquigarrow\, x = y
~~~~~~~~
\dot{\varinfty} = \varinfty
$$
$$
p\,\varinfty \,\land\, p\,0 \,\land\, (\forall x.\,p x \,\rightsquigarrow\, p \dot{x}) \,\rightsquigarrow\, p y
$$
The first axiom ensures the injective property and the second axiom makes the third axiom, the induction principle, work as expected.

Hence $2+2=4$ can be stated in $\Down$ (seeing $+$ as a suitable abbreviation).
It can also be proved, but many other theorems of ordinary mathematics can not be proved, of course
(it does not contain arithmetic in general).

Since $\varphi \vdash \varphi$ we have among others $1 \vdash 1$, but this is just a curiosity.

\subsection{Countable Infinite Indeterminacy}

Let $\omega$ be the axiom:
$$
(\Down x \rightsquigarrow \Down \dot{x}) ~\land~ \exists y.\, \Down y
$$

\medskip

\noindent
No ambiguity is possible with respect to the use of $\omega$ for the set of natural numbers, and the motivation is that
the axiom $\omega$ introduces a countable infinite type in $\Down$.
The first part says that once indeterminate always indeterminate.
The second part of the axiom $\omega$ says that indeterminacy exists.
In other words, we can say that $\omega$ yields $\Down$-\emph{confinement} and $\Down$-\emph{existence}.

With the axiom $\omega$ we extend $\Down$ to the indeterminacy theory $\Downo$
(propositional higher order logic with countable infinite indeterminacy)
such that all theorems of ordinary mathematics can be proved
(the axioms can be shown consistent in axiomatic set theory \cite{Mendelson97}, which is the standard foundation of mathematics
and stronger than $\Downo$, cf.\ the second G\"{o}del incompleteness theorem).

Although the propositional higher order logic $\Down$ is our starting point, the indeterminacy theory $\Downo$ is going to be our most important formal system
and we use $\text{IT}=\{\varphi \mid \omega \vdash \varphi\}$ as a shorthand for its theorems and $\Theta \Vdash \Gamma$ instead of $\Theta,\omega \vdash \Gamma$.
In particular we previously used $\theta \Vdash \varphi$ in the case study (with the conjunction of the formulas {\KB} as the theory $\theta$).

We allow a few more abbreviations:
$$
\hat{\varphi} \,\equiv\, \varphi \land \neg \varphi
~~~~~~~~
\check{\varphi} \,\equiv\, \varphi \lor \neg \varphi
$$
We can now state the interesting property of $\Downo$ (coming from $\Down$) succinctly:
$$
\hat{\varphi} \nVdash \check{\varphi}
$$

\subsection{Classical Logic}

Let $\Up$ be the axiom:
$$
\Up x
$$
The $\Up$ axiom is equivalent to $\Down$-\emph{non-existence}, namely $\nexists x.\, \Down x$, and with the axiom $\Up$ we extend
$\Down$ to the classical propositional higher order logic $\Downc$ which was thoroughly investigated in \cite{Henkin63-FM,Andrews63-FM}.

Finally we can combine the extensions $\Downc$ and $\Downi$ into the classical higher order logic $\Downci$, also known as Q$_0$
based on the typed $\lambda$-calculus, and often seen as a restriction of the transfinite type theory Q \cite{Andrews65}
by removing the transfinite types.
Q$_0$ is implemented in several automated theorem provers with many active users \cite{Gordon85,Paulson94}.
Classical second order logic, first order logic, elementary logic (first order logic without functions and equality)
and propositional logic can be seen as restrictions of Q$_0$.

In contrast to the paraconsistent $\Downo$ the classical $\Downci$ is not a foundation of mathematics,
but we obtain the type theory Q$_0^\sigma$ by replacing the sort $\iota$ with the sort $\sigma$ and adding the relevant Peano postulates
$\dot{x} \neq 0 \,\land\, (x \neq y \rightsquigarrow \dot{x} \neq \dot{y})$ in our notation, cf.\ \cite[pp.~209/217]{Andrews86} for the details.

\subsection{Other Logics}

In order to investigate finite truth tables, namely the three-valued and four-valued logics discussed in previous sections, we have the following abbreviations:
$$
\AN \,\equiv\, \dot{\AF}
~~~~~~
\AP \,\equiv\, \ddot{\AF}
$$
We get the four-valued logic $\Downdd$ by adding the following axiom to $\Down$:
$$
\Up x ~\lor~ x = \AN ~\lor~ x = \AP
$$
Likewise we get the three-valued logic $\Downd$ by adding the following axiom to $\Down$:
$$
\Up x ~\lor~ x = \AN
$$
But here $\AP = \AF$ due to the injection property of the indeterminacy generation.

\section{Logical Semantics of Natural Language}

A paramount application of higher order logic is natural language semantics, in particular in the Montague grammar tradition
of logical semantics \cite{Thomason74,Chierchia+00},
where the grammar and meaning of natural language sentences are defined and the logical consequences of the sentences must in the end be tested against our intuition.
A set of sentences provides a model of the world as observed by a person or, more generally, an agent, and the agent is part of the world as are other agents.
In such cases it is important to be able to reason about the knowledge,
beliefs, assertions and other propositional attitudes of agents.

We think that a robust treatment of propositional attitudes in natural language is critical for many AI applications.
We show in \cite{Villadsen01:LACL} how to obtain a paraconsistent logic for the propositional attitudes
of agents while retaining classical logic for the observer.
The semantics of the natural language sentences can be tested when used in arguments.

The sentences in the arguments below are ambiguous --- depending on the scope of the propositional attitude ---
but we have chosen examples where the ambiguity does not effect the correctness.
For example, the following argument is correct in any case.

\begin{displayproof}
\text{John believes that Victoria smiles and dances.}
\\[~~~$\surd$]
\text{John believes that Victoria smiles.}
\end{displayproof}

\noindent
However, a paraconsistent logic is needed in order to handle the following incorrect argument.

\begin{displayproof}
\text{John believes that Gloria smiles and doesn't smile.}
\\[~~~$\div$]
\text{John believes that Victoria dances.}
\end{displayproof}

\noindent
These sentences and many other can be translated in a rather simple way into formulas of the paraconsistent higher order logic
$\Downo$ presented here with the above correctness and incorrectness results \cite{Villadsen01:LACL}.
We here outline the translation for the following argument, which we consider to be incorrect although
this is of course debatable.

\begin{displayproof}
\text{John believes that Gloria smiles and doesn't smile.}
\\[~~~$\div$]
\text{John believes that Victoria dances or doesn't dance.}
\end{displayproof}

\noindent
The translation is based on a categorial grammar, with a multi-dimensional
type theory as model theory and a sequent calculus as proof theory.
We start with a string, that is, a sequence of so-called tokens, obtained from the sentence in a very simple way.
For instance, the word order is not changed.
$$
\begin{array}{l}
\texttt{John believe Gloria smile and not smile stop}
\\
\texttt{so John believe Victoria dance or not dance stop}
\end{array}
$$
From this string of tokens the categorial grammar provides the following formula
(the details of this rather complex translation can be found in \cite{Villadsen01:LACL}):
$$
\begin{array}{l}
\textsf{so} ~
\\
~~
  (\varlambda \V.\,\textsf{stop} \V ~ \varlambda \V.\,\textsf{believe} \V ~
    (\varlambda \V.\,\textsf{and'} \V ~ \textsf{smile} \V ~ (\textsf{not} \V ~ \textsf{smile} \V) ~
      \textsf{Gloria} \V) ~ \textsf{John} \V) ~
\\
~~
  (\varlambda \V.\,\textsf{stop} \V ~ \varlambda \V.\,\textsf{believe} \V ~
    (\varlambda \V.\,\textsf{or'} \V ~ \textsf{dance} \V ~ (\textsf{not} \V ~ \textsf{dance} \V) ~
      \textsf{Victoria} \V) ~ \textsf{John} \V)
\end{array}
$$

\noindent
The formula contains lexical combinators and a single variable $\V$ ranging over so-called situations (or indices) of type $o$.
To avoid some parentheses we use the convention that the variable $\V$ is always hung on to the preceding term.

The tokens and the combinators are in one-to-many correspondence, which is exemplified by primes \textsf{'} attached to conjunction and disjunction
also available on the sentence level without primes; the sentence level possibility is omitted here for negation however.
As a pleasant property we have that sentences embedded in propositional attitudes are translated independently of the embedding.

The only entities needed for this example are agents (who can have propositional attitudes),
but a richer ontology is of course possible still within type $o$
(we do not count indices as entities, since they are to be thought of as situations).

The lexicon for the present fragment of English has the following systematic abbreviations
(the notation $\bigcirc$ is a place-holder for the constructs listed after $|$).
$$
\begin{array}{l}
\textsf{so} ~\equiv~ \varlambda p q.\,\forall i.\,p i \rightarrow q i
\\[1.5ex]
\textsf{stop} ~\equiv~ \varlambda \V p.\,\ell \V \land p \V
\\[1.5ex]
\textsf{John} ~~ \textsf{Gloria} ~~ \textsf{Victoria} ~\equiv~ \varlambda \V.\bigcirc ~~|~~ J~G~V
\\[1.5ex]
\textsf{know} ~~ \textsf{believe} ~\equiv ~ \varlambda \V p x.\,\forall j.\bigcirc\V x j \rightarrow p j ~~|~~ K~B
\\[1.5ex]
\textsf{and} ~~ \textsf{or} ~\equiv ~ \varlambda \V a b.\,a \bigcirc b ~~|~~ \land~\lor
\\[1.5ex]
\textsf{and'} ~~ \textsf{or'} ~\equiv ~ \varlambda \V t u x.\,t x \bigcirc u x ~~|~~ \land~\lor
\\[1.5ex]
\textsf{not} ~\equiv ~ \varlambda \V t x.\,\neg t x
\\[1.5ex]
\textsf{smile} ~~ \textsf{dance} ~\equiv ~ \varlambda \V x.\bigcirc\V x ~~|~~ S~D
\end{array}
$$

\noindent
Here the variables $a$, $b$, $i$, $j$, $x$ have type $o$ and $p$, $q$, $t$, $u$ have type $oo$.
The last combinators are equal to $S$ and $D$, respectively, but we find that the expanded form emphasizes the pattern.
Notice that we have chosen not to make names dependent on the situation (hence we have the constants $J$, $G$ and $V$ for agents).

In order to obtain a paraconsistent logic for the propositional attitudes of agents while retaining classical logic for the observer we introduce
the following ``integrity'' abbreviation $\ell$ of type $oo$ (using the determinacy predicate $\Up$).
$$
\ell ~\equiv~ \varlambda \V.\,(\forall x j.\,\Up(K \V x j)) \land (\forall x j.\,\Up(B \V x j)) \land (\forall x.\,\Up(S \V x)) \land (\forall x.\,\Up(D \V x))
$$

\noindent
Besides the lexicon some postulates are needed.
Knowledge implies belief, hence we should add the postulate $\forall i x j.\,K i x j \rightarrow B i x j$, and similar postulates
yield properties like introspection for knowledge and belief, cf.\ \cite{Villadsen01:LACL}.

Using the abbreviations in the lexicon the formula given earlier reduces to:
$$
\forall i.\,(\ell i \land \forall j.\,B i J j \rightarrow
                                      S j G \land \lnot S j G) \rightarrow
            (\ell i \land \forall j.\,B i J j \rightarrow
                                      D j V \lor \lnot D j V)
$$
As required the formula does not hold in $\Downo$. The details can be found in \cite{Villadsen01:LACL}.

\section{Conclusion}

We have proposed a paraconsistent higher order logic $\Downo$ with countable infinite indeterminacy and 
have described a case study in the domain of medicine as well as an application in logical semantics of propositional attitudes.

We have presented a sequent calculus for the paraconsistent logic $\Down$ and the simple axiom $\omega$ turning $\Down$ into the many-valued logic $\Downo$.
Another axiom $\Up$ turns $\Down$ into the classical logic $\Downc$.
We would like to emphasize that it is not at all obvious how to get from $\Downc$ to $\Down$
when the usual axiomatics and semantics of $\Downc$ do not deal with the axiom $\Up$ separately as we do here.

Corresponding to the proof-theoretical $\vdash$ we have the model-theoretical $\vDash$ based on the type universes, and we believe that
soundness and completeness results can be obtained (the latter with respect to general models of $\Down$ only).

In a way we try to build a bridge between the HOL and MVL communities. We find both fields to be highly relevant to paraconsistent computational logic.

\end{document}